\documentclass[journal]{IEEEtran}
\usepackage{amsmath,amssymb,amsfonts,amsthm,mathtools}
\usepackage{array}
\usepackage{graphicx}
\usepackage{cite}
\usepackage{booktabs}
\usepackage{xcolor}
\usepackage{dsfont}
\usepackage{tikz}
\usetikzlibrary{positioning, arrows.meta}
\usepackage{graphicx}
\begin{document}

\title{Degeneracy-Aware Resource Allocation for Resilient 6G RAN}

\author{Sayanti~Ghosh,~\IEEEmembership{Member,~IEEE}, Indrakshi~Dey,~\IEEEmembership{Senior~Member,~IEEE}, and Nicola~Marchetti,~\IEEEmembership{Senior~Member,~IEEE}%
\thanks{S.~Ghosh and N.~Marchetti are with the Department of Electrical and Electronic Engineering, Trinity College Dublin, Ireland (saghosh@tcd.ie; nicola.marchetti@tcd.ie). I.~Dey is with the Department of Computing and Mathematics, South East Technological University, Ireland (indrakshi.dey@setu.ie). Supported in part by EU MSCA Project COALESCE under Grant 101130739, US--Ireland R\&D Partnership RI-SFI-23/US/3924, and Research Ireland Grant 13/RC/2077\_P2.}
}

\maketitle

\begin{abstract}
Heterogeneous 6G radio access networks (RANs) must allocate resources reliably under interference, latency limits, imperfect channel state information (CSI), and architectural diversity. We propose a degeneracy-aware resource allocation (DG-RA) framework that casts multi-architecture orchestration as a probabilistic game and, unlike single-solution optimization, deliberately favors allocations realizable by many structurally distinct yet performance-equivalent strategy profiles. Resilience is quantified across three layers through Degeneracy-Weighted Path Robustness (DWPR), Functional Substitution Score (FSS), and an Algorithmic Resilience Quotient (ARQ). Across centralized (C-RAN), open (O-RAN), virtualized (V-RAN), and hybrid RAN architectures, and benchmarked against a fractional-programming optimizer, DG-RA matches the state-of-the-art throughput and outage at the static operating point, then exploits its equivalence set to recover $\sim$$99\%$ of throughput from a resource-unit failure with a single switch, where a single-solution optimizer needs tens of iterations to re-converge. The results recast degeneracy not as a rate booster but as a precomputed resilience reserve for disruption-tolerant 6G orchestration.
\end{abstract}


\section{Introduction}
\IEEEPARstart{B}{5G/6G} radio access networks (RANs) are evolving toward open, virtualized, and intelligent architectures to support ultra-reliable low-latency communications, immersive media, and digital twins \cite{saad2022vision,dang2022what}. Open RAN (O-RAN) exposes programmable control through disaggregated functions and standardized interfaces \cite{eiza2025zerotrust}, centralized RAN (C-RAN) centralizes baseband processing to enable coordinated interference management \cite{park2022cran}, and virtualized RAN (V-RAN) decouples RAN functions from hardware to enable cloud-native flexibility \cite{polese2024empowering}, all consistent with 3GPP NR deployment models \cite{3gpp38211,3gpp38300}. Yet resource allocation across these heterogeneous architectures remains hard because interference coupling, latency budgets, imperfect channel state information (CSI), and architectural diversity interact unpredictably. Optimization, game-theoretic, and learning-based schemes \cite{han2023quantum,luong2019drlsurvey} almost universally drive the system to a \emph{single} preferred operating point, discarding the many structurally distinct allocations that deliver equivalent performance in software-defined RANs.

We argue that this discarded multiplicity is precisely the resource that confers resilience. We capture it through \emph{degeneracy}: the capacity of structurally distinct entities or strategies to deliver equivalent function or performance \cite{Giulio1999}. Unlike redundancy, degeneracy does not replicate structure, which makes it a natural fit for heterogeneous, programmable RANs and a recently recognized lever for robustness in networked systems \cite{dey2025degeneracy}. Exploiting degeneracy lets the network absorb CSI errors, interference swings, and orchestration variability by reallocating across equivalent realizations rather than collapsing onto one fragile optimum.

Building on this insight, we propose a degeneracy-aware resource allocation (DG-RA) framework for multi-architecture 6G RANs with three contributions. (i) A unified, architecture-parameterized model represents C-RAN, O-RAN, V-RAN, and Hybrid RAN within one formulation, and casts orchestration as a probabilistic game whose utility explicitly rewards degeneracy. (ii) Resilience is quantified across the path, function, and algorithm layers via degeneracy-weighted path robustness (DWPR), functional substitution score (FSS), and algorithmic resilience quotient (ARQ) \cite{dey2025degeneracy}, exposing each architecture's dominant resilience layer. (iii) Compared with a fractional-programming optimizer and an interference-aware game, DG-RA achieves the same static operating point while enabling single-step recovery from resource-unit failures, eliminating the need for repeated re-optimization and transforming the equivalence set into a precomputed resilience reserve.

\section{System Model and DG-RA Framework}
\label{sec:system_model}
\begin{figure}[t]
\centering
\resizebox{0.92\columnwidth}{!}{
\begin{tikzpicture}[
    font=\scriptsize,
    >=latex,
    node distance=0.34cm,
    box/.style={draw, rounded corners=3pt, align=center,
        minimum width=6.2cm, minimum height=0.50cm, fill=gray!8, inner sep=2pt},
    metric/.style={draw, rounded corners=3pt, align=center,
        minimum width=1.9cm, minimum height=0.50cm, fill=gray!4, inner sep=1.5pt},
    arrow/.style={->, thick}
]
\node[box] (model) {Unified Architecture-Aware Model\\ C/O/V/Hybrid RAN: power, CSI, interference, latency, QoS};
\node[metric, below left=0.5cm and -1.7cm of model] (dwpr) {DWPR\\(path)};
\node[metric, below=0.5cm of model] (fss) {FSS\\(function)};
\node[metric, below right=0.5cm and -1.7cm of model] (arq) {ARQ\\(algorithm)};
\node[box, below=0.5cm of fss] (utility) {Degeneracy-Aware Utility (rate, latency, power, robustness)};
\node[box, below=of utility] (game) {Probabilistic Game: degeneracy scoring + Gibbs sampling};
\node[box, below=of game] (perf) {Allocation $\rightarrow$ outage $\downarrow$, throughput $\uparrow$, utility $\uparrow$};
\draw[arrow] (model) -- (dwpr); \draw[arrow] (model) -- (fss); \draw[arrow] (model) -- (arq);
\draw[arrow] (dwpr) -- (utility); \draw[arrow] (fss) -- (utility); \draw[arrow] (arq) -- (utility);
\draw[arrow] (utility) -- (game); \draw[arrow] (game) -- (perf);
\end{tikzpicture}
}
\caption{The DG-RA framework: a unified architecture-aware model feeds path-, function-, and algorithm-level robustness into a degeneracy-aware probabilistic game.}
\label{fig:dgra_framework_single}
\end{figure}
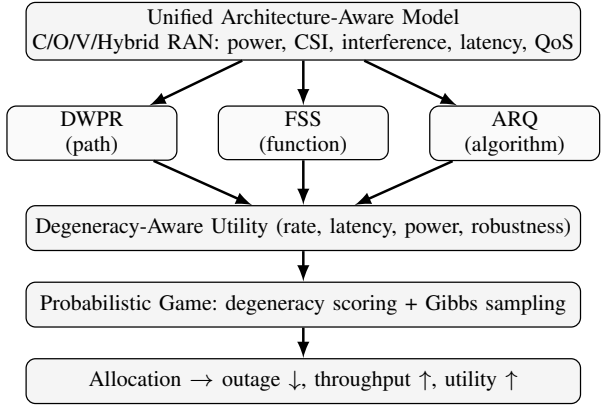

We consider a unified 6G RAN comprising central units (CUs), distributed units (DUs), radio units (RUs), edge/cloud servers, RAN intelligent controllers, and virtual network functions (VNFs) that jointly handle communication, scheduling, orchestration, and interference management. Fig.~\ref{fig:dgra_framework_single} summarizes the framework.

\vspace{-5mm}
\subsection{Unified Architecture Abstraction}
We consider four representative 6G RAN deployment architectures: C-RAN, O-RAN, V-RAN, and Hybrid RAN. Rather than modeling them separately, we adopt a unified parameterized abstraction in which their differences are captured by a small set of deployment-specific parameters while preserving their dominant operational characteristics. Let $\xi\in\{\mathrm{C},\mathrm{O},\mathrm{V},\mathrm{H}\}$ denote C-RAN, O-RAN, V-RAN, and Hybrid RAN, respectively. C-RAN provides centralized baseband processing with strong interference coordination; O-RAN employs disaggregated RAN functions coordinated by the RAN Intelligent Controller (RIC); V-RAN virtualizes baseband processing and deploys VNFs on cloud/edge resources; and hybrid RAN combines centralized control, virtualization, and open RAN intelligence, with core-level optimization and RIC-driven distributed operation.
Each architecture is characterized by
$
\Theta_\xi=
\big(
\chi_\xi^{\mathrm{fh}},
\chi_\xi^{\mathrm{virt}},
\chi_\xi^{\mathrm{ctrl}},
\kappa_\xi,
\omega_{1,\xi},
\omega_{2,\xi},
\omega_{3,\xi}
\big)$,
where $\chi_\xi^{\mathrm{fh}}$, $\chi_\xi^{\mathrm{virt}}$, and $\chi_\xi^{\mathrm{ctrl}}$ are normalized coefficients representing the relative fronthaul, virtualization, and control overheads of architecture $\xi$, with larger values indicating higher overhead (Table~\ref{tab:arch_deg_params}). The parameter $\kappa_\xi>0$ models the interference coupling level, while $\omega_{1,\xi}$, $\omega_{2,\xi}$, and $\omega_{3,\xi}$ weight the contributions of path-, function-, and algorithm-level resilience in the proposed utility function.
\vspace{-5mm}
\subsection{Entities and Degeneracy}
Let $\mathcal{E}=\{e_1,\dots,e_K\}$ be the set of network entities (e.g., RUs, DUs, CUs, VNFs, or edge/cloud servers) and $\mathcal{F}=\{f_1,\dots,f_M\}$ the set of network functions (e.g., scheduling, resource allocation, interference coordination, or packet processing), with $\phi(e_i,f_j)=1$ if entity $e_i$ can realize function $f_j$, and $0$ otherwise. Each entity carries a feature vector $\mathbf{s}_i \in \mathbb{R}^{D}$, where $D$ denotes the feature-space dimension, and the structural dissimilarity between entities $e_i$ and $e_k$ is defined as
$\delta_{i,k}=1-\frac{\mathbf{s}_i^{\!\top}\mathbf{s}_k}{\|\mathbf{s}_i\|\,\|\mathbf{s}_k\|}\in[0,1]$. The degeneracy of $f_j$ is then
\begin{equation}
D(f_j)=\!\!\sum_{\substack{i,k=1\\ i<k}}^{K}\!\!\phi(e_i,f_j)\,\phi(e_k,f_j)\,\delta_{i,k},
\label{eq:degeneracy_re_new}
\end{equation}
which is large when \emph{structurally distinct} entities can perform the same function.
\vspace{-5mm}
\subsection{Physical-Layer and Latency Model}
For OFDM subcarrier set $\mathcal{N}$, serving-RU set $\mathcal{R}_u$ of user $u$, and transmit power allocation $p_i[n]$ (in watts) from RU $i$ on subcarrier $n$, the effective SINR under architecture $\xi$ is
\begin{equation}
\gamma_u^{(\xi)}[n]=
\frac{\sum_{i\in\mathcal{R}_u}|h_{i,u}[n]|^2 p_i[n]}
{\kappa_\xi\sum_{u'\neq u}\sum_{i\in\mathcal{R}_{u'}}|h_{i,u}[n]|^2 p_i[n]+\sigma^2},
\label{eq:sinr_re_new}
\end{equation}
where smaller $\kappa_\xi$ denotes stronger interference coordination. CSI is imperfect: the scheduler observes $\hat{h}_{i,u}[n]=h_{i,u}[n]+e_{i,u}[n]$, $e_{i,u}[n]\sim\mathcal{CN}(0,\sigma_e^2)$, and selects allocations from $\hat{h}$, whereas eq. \eqref{eq:sinr_re_new} is realized on the true $h$. This decision-realization mismatch is the mechanism through which $\sigma_e^2$ degrades reliability. The achievable rate is $R_u^{(\xi)}=\sum_{n\in\mathcal{N}}\log_2(1+\gamma_u^{(\xi)}[n])$ subject to $\sum_{n}p_i[n]\le P_i^{\max}$. End-to-end latency aggregates transmission, processing, and architecture-scaled fronthaul, virtualization, and control delays,
$
L_u^{(\xi)}=L_u^{\mathrm{tx}}+L_u^{\mathrm{proc}}
+\chi_\xi^{\mathrm{fh}}L_u^{\mathrm{fh}}
+\chi_\xi^{\mathrm{virt}}L_u^{\mathrm{virt}}
+\chi_\xi^{\mathrm{ctrl}}L_u^{\mathrm{ctrl}}$,
with QoS constraint $L_u^{(\xi)}\le L_{\mathrm{th}}$.
\vspace{-5mm}
\subsection{Three-Layer Robustness Metrics}
\emph{Path layer.} Let $s$ and $d$ denote the source and destination nodes, respectively, and let $\mathcal{P}_{sd}$ denote the set of all candidate communication paths between them \cite{dey2025degeneracy}. Each path $P_i\in\mathcal{P}_{sd}$ consists of a sequence of communication links. The QoS-compliant path subset is defined as
$
\mathcal{P}_{sd}^{\mathrm{valid}}
=
\left\{
P_i \in \mathcal{P}_{sd} :
\sum_{e\in P_i}\lambda_e \le \Lambda,\;
\min_{e\in P_i}\beta_e \ge \beta_{\min}
\right\}$,
where $\lambda_e$ and $\beta_e$ denote the latency and available bandwidth of link $e$, respectively, while $\Lambda$ and $\beta_{\min}$ are the maximum allowable end-to-end latency and minimum required link bandwidth. DWPR rewards diverse high-quality routes, 
\begin{equation}
\mathrm{DWPR}(s,d)=\frac{1}{|\mathcal{P}_{sd}^{\mathrm{valid}}|}
\!\!\sum_{P_i\in\mathcal{P}_{sd}^{\mathrm{valid}}}\!\!
\mathbf{1}[Q(P_i)\ge\theta]\,\bar{D}_{\mathrm{path}}(P_i),
\label{eq:dwpr_re_new}
\end{equation}
where $Q(P_i)$ is path quality, $\theta$ a threshold, and $\bar{D}_{\mathrm{path}}(P_i)$ the mean diversity of $P_i$ from the remaining valid paths.

\emph{Function layer.} With feasible realization set $\mathcal{E}_{f_j}=\{e_i:\phi(e_i,f_j)=1\}$, the Functional Substitution Score counts structurally distinct substitutes,
\begin{equation}
\mathrm{FSS}(f_j)=\frac{1}{|\mathcal{E}_{f_j}|(|\mathcal{E}_{f_j}|-1)}\sum_{i\neq k}\mathbf{1}[\delta_{i,k}>\delta_f].
\label{eq:fss_re_new}
\end{equation}
where $\delta_f \ge 0$ is the equivalence threshold controlling the similarity required between two function realizations.

\emph{Algorithm layer.} Let $\mathcal{A}_{f_j}$ denote the set of candidate algorithms that can realize function $f_j$, with cardinality $N_A=|\mathcal{A}_{f_j}|$. The ARQ is defined as \cite{dey2025degeneracy},
\begin{equation}
\mathrm{ARQ}(f_j)
=
\frac{1}{N_A(N_A-1)}
\sum_{m\neq\ell}
K_p(m,\ell)\,
D_{\mathrm{alg}}(m,\ell),
\label{eq:arq}
\end{equation}
where $K_p(m,\ell)=\exp\!\left(-\frac{|J_m-J_\ell|}{\sigma_J}\right)$ measures the performance similarity between algorithms $m$ and $\ell$, with $\sigma_J>0$ controlling the sensitivity to utility differences, and
$
D_{\mathrm{alg}}(m,\ell)
=
1-
\frac{\mathbf{a}_m^{\top}\mathbf{a}_\ell}
{\|\mathbf{a}_m\|\,\|\mathbf{a}_\ell\|}
$
denotes their algorithmic diversity, where $\mathbf{a}_m$ and $\mathbf{a}_\ell$ are the corresponding algorithm feature vectors.
\vspace{-5mm}
\subsection{Degeneracy-Aware Utility and Probabilistic Game}
Let
$
\mathbf{U}=(u_1,u_2,\ldots,u_{K_u})\in\mathcal{S}$
denote the joint strategy profile, where $u_k\in\mathcal{S}_k$ is the resource-allocation strategy selected by user (or agent) $k$, and $
\mathcal{S}=\mathcal{S}_1\times\mathcal{S}_2\times\cdots\times\mathcal{S}_{K_u}$
is the corresponding joint strategy space. The mapping
$\mathbf{x}=M(\mathbf{U})$
associates each joint strategy profile with a feasible resource allocation. Since $M(\cdot)$ is non-injective, distinct profiles can yield performance-equivalent allocations; the degeneracy weight of $\mathbf{x}$ is the size of its equivalence class, $w(\mathbf{x})=|\{\mathbf{U}':M(\mathbf{U}')=M(\mathbf{U})\}|$. Exact enumeration is combinatorial, so we employ a tractable estimator. Let
$c\in\mathcal{C}_u$ denote a candidate resource-allocation strategy for user $u$, where $\mathcal{C}_u$ is the candidate strategy set, and let $J_u(c)$ denote the corresponding user utility. The estimated degeneracy weight counts the near-optimal candidate strategies,
$
\hat{w}_u=
\left|
\left\{
c\in\mathcal{C}_u:
J_u(c)\ge
(1-\epsilon)
\max_{c'\in\mathcal{C}_u}J_u(c')
\right\}
\right|$,
where $c'$ is any candidate strategy in $\mathcal{C}_u$ and $\epsilon\in(0,1)$ specifies the width of the near-optimal utility band. This is linear in the candidate-set size and requires no exhaustive search. The per-user utility combines rate, latency, power, and robustness,
\begin{equation}
\begin{aligned}
J_u^{(\xi)}(\mathbf{x})=&\;\alpha_\xi R_u^{(\xi)}-(1-\alpha_\xi)L_u^{(\xi)}-\mu_\xi P_u\\
&+\omega_{1,\xi}\mathrm{DWPR}_u+\omega_{2,\xi}\mathrm{FSS}_u+\omega_{3,\xi}\mathrm{ARQ}_u.
\end{aligned}
\label{eq:utility_re_new}
\end{equation}
Here $\alpha_\xi\in[0,1]$ trades rate against latency and $\mu_\xi\ge0$ is the power price; the latter is denoted $\mu_\xi$ rather than $\lambda_\xi$ to avoid clashing with the link latency $\lambda_e$ and the disruption rate $\lambda$ used in Section~\ref{sec:results}.
To favor allocations supported by many performance-equivalent realizations, we augment the system utility as
$
S^{(\xi)}(\mathbf{x})
=
J^{(\xi)}(\mathbf{x})
+
\eta_\xi
\log\!\left(1+\hat{w}(\mathbf{x})\right)$, where $J^{(\xi)}(\mathbf{x})=\sum_{u=1}^{K_u}J_u^{(\xi)}(\mathbf{x})$ is the aggregate system utility, $\hat{w}(\mathbf{x})$ is the estimated degeneracy weight, and $\eta_\xi\ge0$ controls the influence of the degeneracy term. The logarithmic function provides diminishing returns, and resource selection follows the Gibbs distribution
\begin{equation}
\mathbb{P}^{(\xi)}(\mathbf{x})=
\frac{\exp(S^{(\xi)}(\mathbf{x})/T)}
{\sum_{\mathbf{x}'\in\mathcal{X}}\exp(S^{(\xi)}(\mathbf{x}')/T)},
\label{eq:allocation_distribution_re_new}
\end{equation}
with temperature $T$ over the feasible allocation space $\mathcal{X}$. Setting $\eta_\xi=0$ recovers a degeneracy-agnostic interference-aware best-response (IA-BR) baseline, which converges to a single equilibrium and discards the equivalence set. The system is solved by damped best response: a profile $\mathbf{U}^\star$ is an equilibrium when no user can raise its own $J_u^{(\xi)}$ by a unilateral deviation. Degeneracy makes this equilibrium non-unique. We define the \emph{degenerate equilibrium set} as the equilibria that share the optimal payoff,
\begin{equation}
\mathcal{E}_{\xi}^{\star}=
\big\{\mathbf{U}^\star\!:\mathbf{U}^\star \text{ is an equilibrium},\ J_u^{(\xi)}(M(\mathbf{U}^\star))=J_u^\star\ \forall u\big\},
\label{eq:pne_set_re_new}
\end{equation}
where $J_u^\star$ is the equilibrium payoff of user $u$. Eq. \eqref{eq:pne_set_re_new} is the intersection of the no-deviation condition with a payoff level set, so its members are interchangeable at runtime without utility loss; its cardinality is the operational embodiment of degeneracy.
\vspace{-5mm}
\subsection{Architecture Instantiation}
\label{sec:architectural_cases}
The dominant source of degeneracy differs by architecture: centralized scheduling makes path robustness (DWPR) dominant in C-RAN; multiple VNF placements make functional substitution (FSS) dominant in V-RAN; RIC-driven control spreads degeneracy across all layers in O-RAN; and Hybrid RAN inherits all three, giving the richest strategy space. Table~\ref{tab:arch_deg_params} summarizes the architectural taxonomy, where the normalized coefficients are classified as Low, Moderate, or High relative to a normalized reference architecture (unit overhead) to enable qualitative comparison across the considered RAN architectures. The corresponding resilience characteristics are validated in Section~\ref{sec:results}.

\begin{table}[b]
\centering
\caption{Architecture-Specific Degeneracy Characteristics}
\label{tab:arch_deg_params}
\scriptsize
\setlength{\tabcolsep}{3pt}
\renewcommand{\arraystretch}{1.1}
\begin{tabular}{p{1.05cm} p{0.6cm} p{0.7cm} p{1.25cm} p{2.45cm}}
\toprule
\textbf{Arch.} & $\chi^{\mathrm{fh}}$ & $\chi^{\mathrm{virt}}$ & \textbf{Dominant} & \textbf{Main degeneracy source} \\
\midrule
C-RAN  & High & Low  & DWPR & Centralized scheduling/power \\
O-RAN  & Mod. & Mod. & DWPR, FSS, ARQ & Distributed open orchestration \\
V-RAN  & \shortstack{Low/\\Mod.} & High & FSS & VNF placement and mapping \\
Hybrid & High & High & DWPR, FSS, ARQ & Joint coordination + virtualization \\
\bottomrule
\end{tabular}
\end{table}

\section{Performance Analysis}
\label{sec:performance}
Because $M(\cdot)$ is non-injective, the outage of user $u$ averages over the allocation distribution, $P_{\mathrm{out},u}^{(\xi)}=\sum_{\mathbf{x}}\mathbb{P}^{(\xi)}(\mathbf{x})\,\mathbb{P}(\gamma_u^{(\xi)}(\mathbf{x})<\gamma_{\mathrm{th}})$. Writing the desired and residual-interference powers of eq. \eqref{eq:sinr_re_new} as $S_u$ and $I_u$, this is $\mathbb{P}(S_u<\gamma_{\mathrm{th}}(\kappa_\xi I_u+\sigma^2))$. Crucially, the expression retains interference: as the transmit-power budget is scaled by a factor $\rho>0$, both the desired signal power $S_u$ and the residual interference power $I_u$ scale proportionally, yielding $\gamma_u^{(\xi)}\!\to\!S_u/(\kappa_\xi I_u)$. Hence,
\begin{equation}
P_{\mathrm{out},u}^{(\xi),\infty}
=
\mathbb{P}\!\left(
\frac{S_u}{\kappa_\xi I_u}
<
\gamma_{\mathrm{th}}
\right),
\label{eq:outage_floor_re_new}
\end{equation}
which represents an irreducible high-SNR outage floor under interference-limited operation, since the limiting SINR depends only on the ratio $S_u/(\kappa_\xi I_u)$. Imperfect CSI further raises the outage floor by causing suboptimal resource allocation and increased residual interference. Any interference-aware allocation, such as DG-RA and FP (Fractional Programming) alike, drives $I_u$ down to reach this floor; what distinguishes DG-RA is that $\mathbb{P}^{(\xi)}(\mathbf{x})$ concentrates on a \emph{set} of such low-$I_u$, structurally distinct allocations rather than a single one, so an equally good operating point remains available when part of the network is lost. The ergodic throughput
$\bar{R}_u^{(\xi)}=\sum_{\mathbf{x}}\mathbb{P}^{(\xi)}(\mathbf{x})\mathbb{E}_{h}[\log_2(1+\gamma_u^{(\xi)})]$
yields the system throughput
$R_{\mathrm{sum}}^{(\xi)}=\sum_{u=1}^{K_u}\bar{R}_u^{(\xi)}$,
where $K_u$ is the number of users. Both depend on the allocation distribution $\mathbb{P}^{(\xi)}(\mathbf{x})$, which favors allocations with larger $\hat{w}(\mathbf{x})$ by averaging performance over multiple performance-equivalent operating points rather than a single allocation.
\section{Results and Discussion}
\label{sec:results}
We evaluate a multi-user orthogonal frequency division multiple access system under Rayleigh fading with imperfect CSI via Monte Carlo simulation across all four architectures with the settings in Table~\ref{tab:system_parameters}. We compare DG-RA against three baselines: \emph{FP}, a fractional-programming sum-rate optimizer (quadratic transform, gradient projection) representing the state of the art; \emph{IA-BR}, an interference-aware best-response game (the same Nash operating point DG-RA targets, but without maintaining the equivalence set); and \emph{MaxSINR}, a naive single-strongest-RU heuristic. Each operating point is averaged over 300 independent Monte Carlo realizations, and convergence is observed over 160 iterations. The headline message is deliberate: DG-RA does not beat the state of the art at the static operating point, it \emph{matches} it, and its value appears under disruption. We report six observations.
\begin{figure*}[t]
\centering
\includegraphics[width=0.99\textwidth]{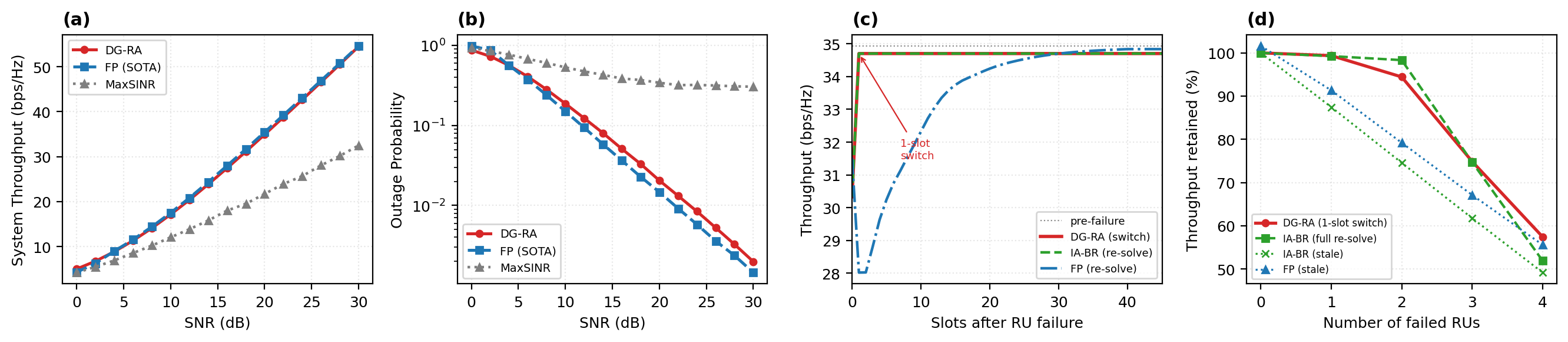}
\caption{Representative results for the Hybrid RAN architecture. DG-RA versus a state-of-the-art optimizer (FP) and baselines. (a) Throughput and (b) outage versus SNR: DG-RA tracks FP almost exactly and both leave the naive MaxSINR far behind, so degeneracy costs nothing at the static operating point. (c) Recovery trajectory after a random RU failure: DG-RA restores full throughput in a single slot by switching within its equivalence set, IA-BR re-converges in a few best-response slots, and FP crawls back over $\sim$$40$ gradient slots. (d) Graceful degradation: a one-slot DG-RA switch retains as much throughput as a full IA-BR re-solve and stays well above the stale point-optimizers as more RUs fail.}
\label{fig:perf}
\end{figure*}
\begin{table}[b]
\centering
\caption{Simulation Parameters (5G NR-Based) \cite{3gpp38211,3gpp38300}.}
\label{tab:system_parameters}
\footnotesize
\setlength{\tabcolsep}{4pt}
\renewcommand{\arraystretch}{1.02}
\begin{tabular}{ll}
\toprule
\textbf{Parameter} & \textbf{Value} \\
\midrule
Carrier freq.\ $f_c$ / bandwidth $B$ & 3.5 GHz / 20 MHz \\
Subcarriers $N$ / spacing & 64 / 15 kHz \\
Users $K_u$ / resource units & 6 / 8 \\
Noise Power Spectral Density & $-174$ dBm/Hz \\
SNR range / SINR threshold $\gamma_{\mathrm{th}}$ & 0--30 dB / 3 dB \\
CSI error variance $\sigma_e^2$ & 0.03--0.11 \\
Monte Carlo trials / $\epsilon$-band & 300 / 0.06 \\
Coupling $\kappa_\xi$ (C/O/V/H) & 0.45 / 0.65 / 0.85 / 0.40 \\
Trade-off $\alpha_\xi$ (C/O/V/H) & 0.70 / 0.65 / 0.60 / 0.68 \\
$(\omega_1,\omega_2,\omega_3)_\xi$: C; O & .26,.10,.12; .20,.18,.10 \\
\hphantom{$(\omega_1,\omega_2,\omega_3)$} V; H & .14,.24,.08; .24,.20,.12 \\
\bottomrule
\end{tabular}
\end{table}
The physical-layer parameters follow the 5G NR specifications in 3GPP TS 38.211 \cite{3gpp38211} and TS 38.300 \cite{3gpp38300}. The architecture-dependent coefficients $(\kappa_\xi,\alpha_\xi,\boldsymbol{\omega}_\xi)$ are normalized design parameters, where $\boldsymbol{\omega}_\xi=(\omega_{1,\xi},\omega_{2,\xi},\omega_{3,\xi})$ denotes the architecture-specific resilience weights for DWPR, FSS, and ARQ, respectively, as summarized in Table~\ref{tab:system_parameters}.

\textbf{1) Static operating point: parity with the state of the art.}
Fig.~\ref{fig:perf}(a)--(b) establish the baseline fact on which everything else rests: at the static operating point DG-RA is indistinguishable from FP. At 30 dB DG-RA reaches $54.5$ bps/Hz against FP's $54.6$, and an outage of $2.0\times10^{-3}$ against FP's $1.5\times10^{-3}$, while the naive MaxSINR stalls at $32.5$ bps/Hz and an interference floor near $3\times10^{-1}$. The interpretation matters: because DG-RA targets the same interference-aware Nash point as IA-BR and FP, the degeneracy machinery imposes \emph{no} rate or reliability penalty. The honest claim is parity, not superiority; the advantage is not visible until the network is perturbed.

\textbf{2) Recovery trajectory: $O(1)$ versus re-optimization.}
Fig.~\ref{fig:perf}(c) introduces a random RU failure after allocation and tracks throughput evolution slot by slot. Following the failure, all schemes exhibit an initial one-slot drop due to the invalidation of the pre-failure allocation, before adapting to the new network state. DG-RA then holds the performance-equivalence set eq. \eqref{eq:pne_set_re_new}, so a single table lookup switches the affected users to a surviving equivalent and throughput is restored in \emph{one} slot. The single-solution optimizers, i.e., IA-BR and FP, must instead recompute a feasible allocation after the failure: IA-BR needs several best-response iterations to reach a new equilibrium, and FP re-solves the optimization and returns to the operating point only after $\sim$40 gradient iterations, during which degraded service is delivered. The mechanism is exactly the one the framework was built for: redundancy without replication, realized as structurally distinct fallbacks that are already QoS-equivalent, turning re-optimization latency into a precomputed lookup.

\textbf{3) Graceful degradation under multiple failures.}
Fig.~\ref{fig:perf}(d) sweeps the number of simultaneously failed RUs and reports the throughput retained one slot after the failure. A single DG-RA switch, i.e., replacing the failed allocation with a precomputed performance-equivalent allocation from the degeneracy set, retains $99\%$ throughput after one failure and degrades gracefully to $94\%$ and $75\%$ after two and three failures, respectively. This closely matches a \emph{full} IA-BR re-solve while requiring only a single lookup; in contrast, the single-solution optimizers retain only $87\%$, $75\%$, and $62\%$ over the same range. The curve also exposes the limit of the mechanism: once enough RUs are lost that the equivalence set is exhausted (four of eight), all schemes converge near $50$--$57\%$, because there is no longer a structurally distinct fallback to switch to. Degeneracy enables graceful degradation only while the equivalence set remains populated, a property quantified by the robustness metrics in Observation~6.

\begin{figure}[t]
\centering
\includegraphics[width=0.65\columnwidth]{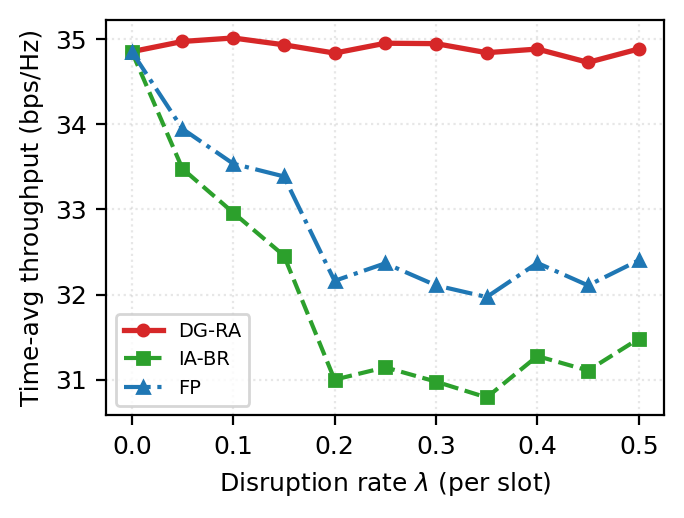}
\caption{Disruption-rate robustness. For the Hybrid RAN architecture, under a stream of failures at rate $\lambda$, DG-RA's time-averaged throughput is flat because each disruption is absorbed by a one-slot switch, while the point-optimizers lose ground as they spend more time re-solving.}
\label{fig:drift}
\end{figure}

\textbf{4) Disruption-rate robustness: a dynamic-network advantage.}
Fig.~\ref{fig:drift} stresses the network with a stream of disruptions at rate $\lambda$, charging each point-optimizer a re-solve latency during which it runs a stale allocation. DG-RA's time-averaged throughput remains nearly constant at $\sim$34.9~bps/Hz across the entire range because each disruption is absorbed by a one-slot switch to a performance-equivalent allocation. In contrast, IA-BR and FP decrease to 31.5 and 32.4~bps/Hz, respectively, as disruptions become more frequent and an increasing fraction of time is spent recomputing feasible allocations. The static operating points are identical (Observation~1); the separation here is created entirely by recovery latency, which is precisely the quantity degeneracy removes. The advantage therefore grows with the dynamism of the environment, the regime that matters most for 6G orchestration.

\begin{figure}[t]
\centering
\includegraphics[width=0.65\columnwidth]{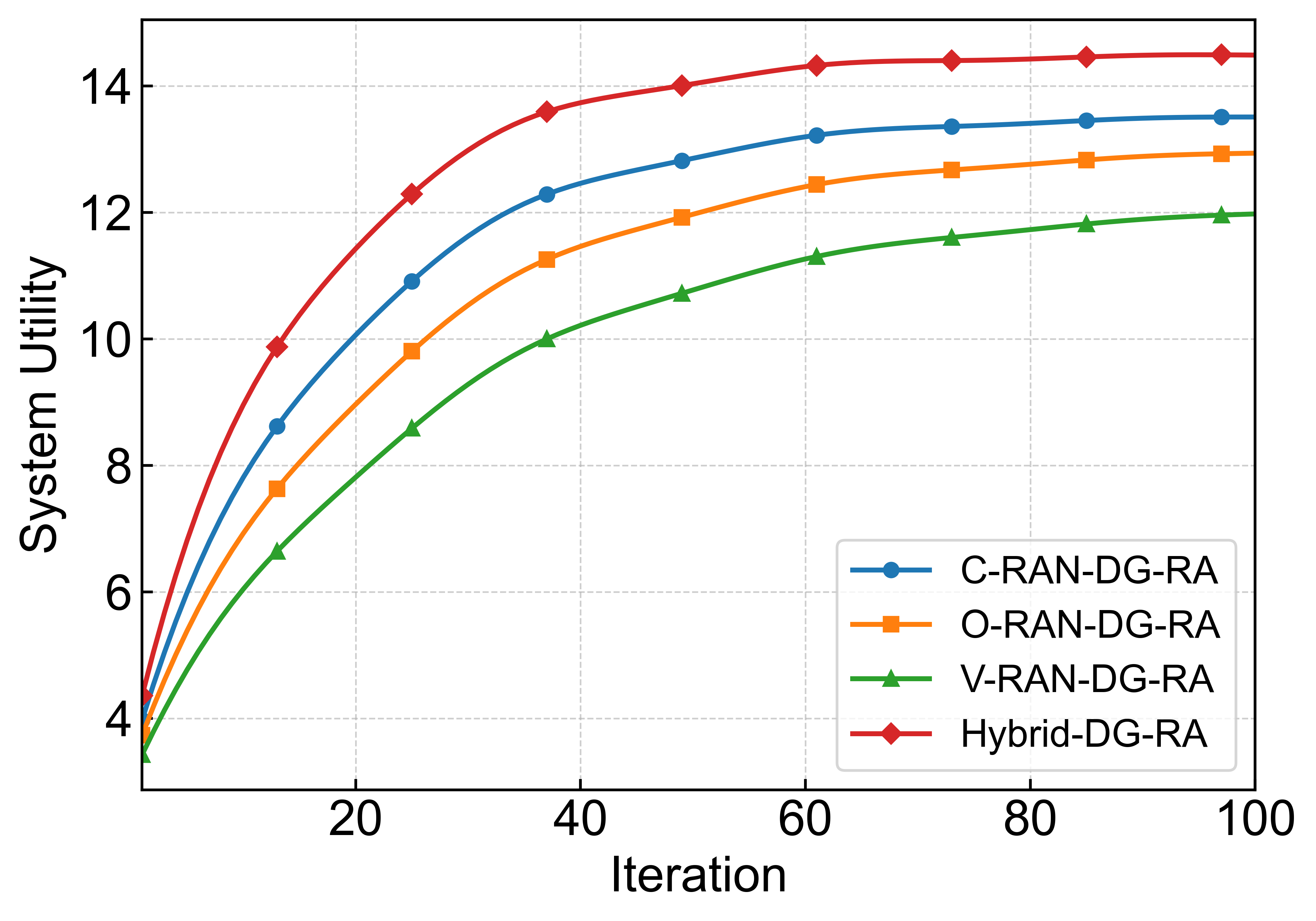}
\caption{Convergence of system utility over iterations. DG-RA converges to stable high-utility operating regions supported by multiple equivalent strategies.}
\label{fig:utility_convergence}
\end{figure}

\textbf{5) Convergence: settling into a set, not a point.}
Fig.~\ref{fig:utility_convergence} shows the ensemble-averaged degeneracy-aware score $\mathbb{E}[S^{(\xi)}(\mathbf{x}^{(t)})]$, where $t$ denotes the iteration index. Starting from a random high-interference allocation, DG-RA converges monotonically to a stable operating point without oscillation. Stability arises from the degeneracy term $\log(1+\hat{w}(\mathbf{x}))$, which broadens the high-score basin and causes the best response in eq. \eqref{eq:allocation_distribution_re_new} to relax toward the degenerate equilibrium set eq. \eqref{eq:pne_set_re_new}, rather than converging to a single fragile optimum, resulting in smooth convergence. Convergence \emph{speed} is a measurable proxy for degeneracy richness: Hybrid reaches $90\%$ of its final utility in $15$ iterations, C-RAN in $21$, O-RAN in $28$, and V-RAN in $41$, and the ordering tracks the available equivalence-set size, which is also the recovery capacity discussed in Observation~2.

\begin{figure}[t]
\centering
\includegraphics[width=0.70\columnwidth]{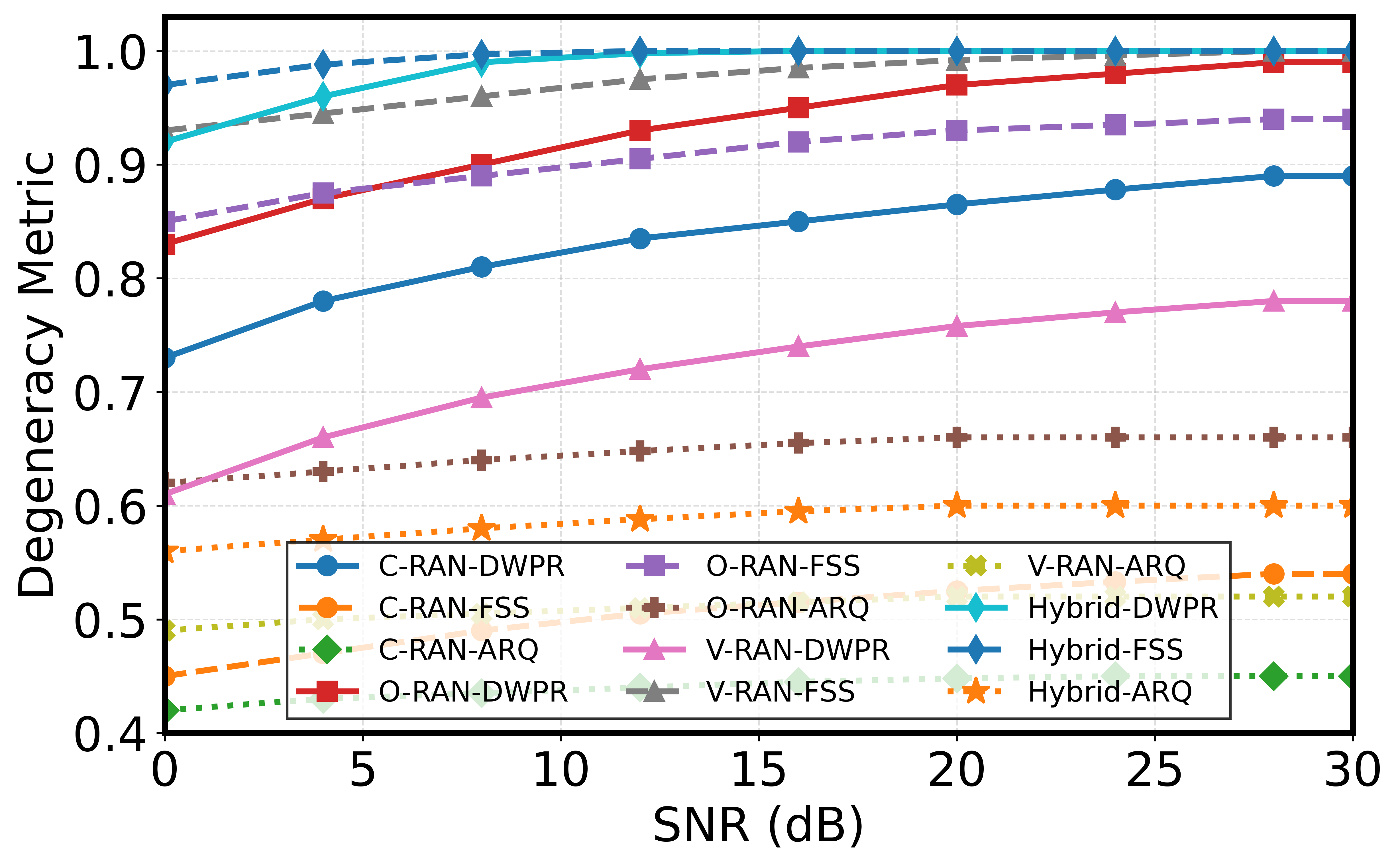}
\caption{DWPR, FSS, and ARQ versus SNR for C-RAN, O-RAN, V-RAN, and Hybrid RAN. DWPR increases with SNR, whereas FSS and ARQ remain largely architecture-dependent with only weak SNR variation.}
\label{fig:dwpr_fss_snr}
\end{figure}

\textbf{6) Robustness metrics: a per-architecture resilience fingerprint.}
Fig.~\ref{fig:dwpr_fss_snr} illustrates the three degeneracy metrics that characterize the size and diversity of the performance-equivalent recovery set for each architecture. DWPR increases with SNR because improved channel quality enlarges the set of QoS-compliant recovery paths in eq. \eqref{eq:dwpr_re_new}, whereas FSS and ARQ exhibit only weak SNR dependence since they are primarily determined by function and algorithm diversity, respectively. Consequently, each architecture exhibits a distinct degeneracy fingerprint: C-RAN is path-dominated, V-RAN is function-dominated, O-RAN provides balanced path, function, and algorithm diversity through the RIC, and Hybrid maintains consistently high degeneracy across all three dimensions. These fingerprints identify the dominant source of recovery alternatives and agree with the taxonomy in Table~\ref{tab:arch_deg_params}.

\section{Conclusion}
\label{sec:conclusion}
We presented DG-RA, a degeneracy-aware probabilistic game-theoretic resource-allocation framework that unifies C-RAN, O-RAN, V-RAN, and Hybrid RAN. Against a fractional-programming optimizer and an interference-aware game, DG-RA matches the state-of-the-art operating point in both throughput and outage while imposing no static penalty; its contribution is resilience, not peak rate. By maintaining a set of performance-equivalent profiles, DG-RA recovers full throughput from a resource-unit failure in a single slot, where a single-solution optimizer needs a few to tens of solver iterations to re-converge; DG-RA degrades gracefully as failures accumulate and, under a stream of disruptions, holds a flat time-averaged throughput while the optimizers lose ground to re-solve latency. The DWPR--FSS--ARQ fingerprint summarizes each architecture's recovery capability through alternative paths, function realizations, and algorithms. Higher values indicate larger performance-equivalent recovery sets and greater resilience. Degeneracy thus serves as a precomputed resilience reserve, enabling one-slot recovery via performance-equivalent allocations. Future work will consider correlated failures, user churn, and online adaptation of the architecture weights.

\bibliographystyle{IEEEtran}
\bibliography{reference}
\end{document}